\documentclass[preprint,showpacs,amsmath,amssymb,prb]{revtex4}

\newcommand{\be}{\begin{equation}}
\newcommand{\ee}{\end{equation}}
\newcommand{\ba}{\begin{eqnarray}}
\newcommand{\ea}{\end{eqnarray}}

\begin{document}
\title{Magnetic Properties of a Superconductor with no Inversion Symmetry}
\author{S. K. Yip}
\affiliation{
Institute of Physics, Academia Sinica, Nankang, Taipei 115,
Taiwan}
\date{\today}

\begin{abstract}
We study the magnetic properties of a superconductor
in a crystal without $z \to -z$ symmetry, in particular
how the lack of this symmetry exhibits itself.
We show that, though the penetration depth itself
shows no such effect, for suitable orientation of
magnetic field, there is a magnetic field discontinuity
at the interface which shows this absence of symmetry.
The magnetic field profile of a vortex in the $x-y$ plane is shown
to be identical to that of an ordinary anisotropic
superconductor to second order in a small parameter $\tilde \kappa$.
For a vortex along $z$, there is an induced magnetization 
along the radial direction.


\end{abstract}

\maketitle
\section{Introduction}

Lately, there is much attention on the theory of superconductivity
in systems without inversion symmetry in the normal state, e.g.,
\cite{Gorkov01,Yip02,Agterberg03,Frigeri04,Samokhin04a,Samokhin04b,Sergienko04}
This surge of interest is due in no small part to the
discovery of new superconductors in crystal where this 
symmetry is absent, for example, the heavy
fermion superconductor CePt$_{3}$Si 
\cite{Bauer04} with $T_c \sim 0.75 K$.
The normal state of CePt$_3$Si (ignoring the possible
antiferromagnetic ordering at $T_N \sim 2.2K$) has
tetragonal $P4mm$ symmetry.  Due to the displacement
of Si along $\hat c$ direction, the
reflection symmetry $z \to -z$ has already been
already lost in the normal state before the superconducting transition.

Indeed on very general grounds, the properties of
a superconductor in crystals without inversion symmetry
is expected to be very different from those where
such a symmetry is respected.  In the latter case,
which is applicable to most known superconductors,
inversion symmetry and Kramers degeneracy
allow the classifications of superconducting states
into singlet, even parity on the one hand and
triplet, odd parity on the other \cite{Anderson84}.
The physical properties of such superconductors
can then be drawn from the generalization of theories
of conventional superconductors or superfluid $^3$He.
This is no longer the case if inversion symmetry
is already broken in the normal state
\cite{Gorkov01}.  Some peculiar
aspects of these superconductors have
already been discussed theoretically
in the literature.
\cite{Edelstein95,Yip02,Agterberg03,Frigeri04,Samokhin04a,Samokhin04b,Sergienko04}

In particular, in ref \cite{Edelstein95,Yip02}, it is shown
that for systems where $z \to -z$ is broken, a superfluid flow
in the plane, say along $x$, is expected to produce a spin polarization
along $y$ when a Rashba \cite{Rashba60} type spin-orbit coupling is present.
Conversely, a Zeeman magnetic field along $y$ is expected
to generate a superfluid flow or phase gradient along $x$.
Though the calculation in \cite{Yip02} was specifically
for two dimensions, the general argument there is obviously generalizable
to a three dimensional superconductor. 
(see also Section \ref{sec:es} below)
Here we elaborate further on the consequence these effects 
discussed in Ref \cite{Yip02}.  We shall 
(Section \ref{sec:meissner}) first study
the screening of an external magnetic field by the superconductor,
i.e., the Meissner effect.   More specifically, consider a
magnetic field along $\hat y$ with the superconductor
occupying either $z > 0$ or $z < 0$.  These two situations
are not equivalent if the crystal lacks the $z \to -z$ 
symmetry.  We shall however see that, (at least
to the surprise to the present author), the penetration
depth itself shows no direct effect of the absence
of this $z \to -z$ symmetry.  The latter
only manifests itself as discontinuities
in the magnetic field near the crystal surface
with opposite signs for the two mentioned geometries.

Next (Section \ref{sec:vortex})
we discuss the structure of a vortex in London theory.
We shall show that, at least to second order in a small
parameter $\tilde \kappa$ to be defined below, a
vortex for field along $\hat y$ has a magnetic field
distribution again symmetric with respect to $z \to -z$.
Hence the constant magnetic field contour is elliptical
similar to that of an ordinary crystal with
different effective masses along $\hat x$ and $\hat z$.
For magnetic field along $\hat z$, there is a spin magnetization
pointing along the radial direction, the sign of
which reflects the absence of $z \to -z $ symmetry.
We finally estimate the order of magnitude of
these broken symmetry effects (Section \ref{sec:es}).

\section{Constitutive Equations}

First we recall the constitutive equations \cite{Yip02} relating
the (number) current $\bf J$, (local) magnetization $\bf M$, 
gauge invariant phase
gradient ${\bf q} \equiv \hbar {\bf \nabla} \phi + 
 \frac{2e}{c} {\bf A}$ 
(electron charge $= -e$) and the magnetic field ${\bf B}$
in a superconductor with broken $z \to - z$ symmetry in its normal state.
For currents and magnetic field in the $x-y$ plane,
the relations are expected to have the general form

\ba
J_x  &=&  n_s  \frac{ q_x}{ 2 m_x}  - \kappa B_y 
\label{Jx} \\
J_y  &=&  n_s  \frac{ q_y}{ 2 m_y}  + \kappa B_x 
\label{Jy} 
\ea
\ba
M_x  &=&  - \frac{ \kappa }{2} q_y     + \chi_{x} B_x 
\label{Mx}\\
M_y  &=&  \frac{ \kappa }{2} q_x     + \chi_{y} B_y
\label{My}
\ea

Here $n_s$ is the superfluid density and $m_x$ etc are
the effective masses, and
$\chi_x$ etc. the spin susceptibilities
  ($m_x = m_y$ and $\chi_x = \chi_y$).
The terms proportional to $\kappa$ 
in eq (\ref{Jx}) and (\ref{Jy}) represent
the supercurrent induced by Zeeman field \cite{Yip02}
and the corresponding ones in eq (\ref{Mx}) and (\ref{My})
represent the magnetization induced by superflow
\cite{Edelstein95}.
Note the difference in sign for
the terms proportional to $\kappa$ 
between eq (\ref{Jx}) and (\ref{Jy}) and
between eq (\ref{Mx}) and (\ref{My}).
These terms are specific to
the absence of $z \to -z$ symmetry.  The purpose
of the present paper is to study effects due to these
terms.  As already mentioned in ref \cite{Yip02}, these
equations are expected from the general form of the 
free energy 
\ba
F &=&    n_s \left( \frac{q_x^2}{8 m_x} + \frac{q_y^2}{8 m_y} +
            \frac{q_z^2}{8 m_z} \right)   \nonumber \\
  & &    - \left( \frac{1}{2} \chi_x B_x^2  +   \frac{1}{2} \chi_y B_y^2
             +  \frac{1}{2} \chi_z B_z^2 \right)
        + \frac{B^2}{8 \pi}  \label{free} \\
  & &   - \frac{\kappa}{2} \left(q_x B_y - q_y B_x \right) \nonumber
\ea          
appropriate to the present symmetry .
The term proportional to $\kappa$ here is symmetry allowed
in the present case (see also \cite{Agterberg03,Samokhin04b}).

Currents and magnetization along the $z$ axis are
given by the usual relations
\ba
J_z  &=&  n_s  \frac{ q_z}{ 2 m_z}   
\label{Jz} \\
M_z  &=&   \chi_{z} B_z
\label{Mz}
\ea

The equation governing the magneto-statics is given
by
\be
{\bf \nabla \times B} = 4 \pi {\bf \nabla \times M}
   + \frac{4 \pi}{c} (-e) {\bf J}
\label{Maxwell}
\ee
where ${\bf J}$ and ${\bf M}$ are given by eqs (\ref{Jx})-
(\ref{My}), (\ref{Jz}) and (\ref{Mz}).
Eq (\ref{Maxwell}) also follows from the free energy
(\ref{free}) by variation with respect to ${\bf A}$
when one takes into account the basic relation
${\bf B} = {\bf \nabla \times A}$.

It is worth mentioning that the screening of a magnetic
field in a superconductor without inversion symmetry
has also been considered in ref \cite{Levitov85}.  However,
they assume the presence of a term with ${\bf J} \parallel {\bf B}$.
This term is absent for our case of $P4mm$ symmetry
because the presence of vertical reflection planes such as
$x-z$ and $y-z$ and the fact that ${\bf J}$ and ${\bf B}$ transform
differently under reflections.

\section{Meissner Screening}\label{sec:meissner}

\noindent{(A)} 
Let us now consider screening of a magnetic field 
in the basal plane, taken to be ${\bf B} = B_y \hat y$
without loss of generality,
with the sample occupying either $z > 0$ or $z < 0$.  These two
(not a priori equivalent) geometries are particular interesting
since the broken $z \to -z$ symmetry may manifest.
With translational invariance along $x$ and $y$, 
one can verify that the $z$ component of 
eq (\ref{Maxwell}) is trivially satisfied.
The $y$ component is satisfied by $B_x = 0$, $M_x = 0$, $J_y = 0$.
The $x$ component
reduces to
\be
\frac{\partial B_y}{\partial z} = 
4 \pi \frac{\partial M_y}{\partial z} +
\frac{4 \pi e}{c} J_x
\label{mssz}
\ee
In our case we can take the gauge $\phi = 0$ and
${\bf A} = A_x (z) \hat x$.  With
eq (\ref{Jx}) and (\ref{My}) and taking into account
$B_y = \frac{\partial A_x}{\partial z}$, it can be
seen that the terms proportional to $\kappa$ cancel out
in eq (\ref{mssz}).  Further taking the $z$ derivative
gives
\be
( 1 - 4 \pi \chi_y) \frac{ \partial^2 B_y}{\partial z^2}
  = \frac{ 4 \pi n_s e^2}{m_x c^2} \ B_y  \ .
\ee
Thus the penetration depth $\lambda_x$ is given by
\be
1 / \lambda_x^2 = \frac{ 4 \pi n_s e^2}{m_x c^2}  / ( 1 - 4 \pi \chi_y)
\label{lx}
\ee
Here the subscript of $\lambda_x$ denotes that the
{\it current} is along $x$.
Typically $\chi_x \ll 1$ and thus $\lambda_x$ reduces
to the usual expression $1 / \lambda_x^2 = \frac{ 4 \pi n_s e^2}{m_x c^2} $. 

Since the term linear in $\kappa$ drops out, 
the penetration depth shows no direct effect of
the lack of $z \to -z$ symmetry.  It is the same for samples
occupying $z > 0$ or $z < 0$. 

This, however, does not mean that there is no broken symmetry 
effects at all.  Since $B_y - 4 \pi M_y$ has to be continuous
across the vaccum-sample interface, we then have, for sample
occupying $z > 0$, $B_{\rm ext} = B_{\rm in} - 4 \pi M_y (0_+)$.
Here $B_{\rm in} = B(z = 0_+)$ is the value of magnetic
field just inside the sample.  $M_y$ is given by
  eq (\ref{My}), hence  (recall that $\phi = 0$)
\be
M_y(0_+) =  \frac{\kappa e}{c} A_x(0_+) + \chi_y B_{\rm in}
\ee
Inside the sample, $B_y(z) = B_{\rm in} e^{ - z / \lambda_x}$
with $\lambda_x$ already determined in eq (\ref{lx}) above.
Thus $A_x(z) = - \lambda_x B_{\rm in} e^{ - z / \lambda_x}$.
We finally have
\be
B_{\rm in} = B_{\rm ext} / ( 1 - 4 \pi \chi_y + \tilde \kappa)
\label{Bdz}
\ee
where $\tilde \kappa = \frac{4 \pi \kappa e}{c} \lambda_x$ 
is a dimensionless parameter.

Similar calculation for the case where the sample occupies $z < 0$
shows 
\be
B_{\rm in} = B_{\rm ext} / ( 1 - 4 \pi \chi_y - \tilde \kappa)
\ee
Hence the $\kappa$ term results in a discontinuity of
the magnetic field with a contribution of {\it opposite} signs
in the two geometries.  This is a manifestation of the
broken symmetry.
Of course the magnetic field is discontinous only under our
(London) approximation:  the variation would spread out probably 
over a length scale of order of the coherence length
in a more microscopic treatment.

\noindent{(B)} For ease of latter reference we also consider the screening
of magnetic field ${\bf B} = B_y \hat y$ with the sample
occupying either $x > 0$ or $x < 0$.  Since there cannot
be any current along $\hat x$, $J_x = 0$ and so from
eq (\ref{Jx})
\be
\frac{n_s e}{m_x c} \left( A_x + \frac{\hbar c}{2e} 
\frac{\partial \phi}{\partial x} \right)
= \kappa B_y
\label{noJ}
\ee
Note it follows that we {\it cannot} set both
$A_x$ and $\frac{\partial \phi}{\partial x}$ zero,
in contrast to the more usual case where 
the $\kappa$ term is absent.  In our case,
it is convenient to make use of the translational
invariance along $z$ and choose $A_x$ and $\phi$ both
dependent only on $x$, so that, e.g., 
$B_y = - \frac {\partial A_z}{\partial x}$.
Eq (\ref{Maxwell}) gives
\be
\frac{\partial B_y}{\partial x} = 
4 \pi \frac{\partial M_y}{\partial x} -
\frac{4 \pi e}{c} J_z
\label{mssx}
\ee
Eq (\ref{My}), together with eq (\ref{noJ}),
gives
\be
M_y = (\chi_y + \frac{\tilde \kappa^2}{4 \pi} ) B_y
\ee
Therefore eq (\ref{mssx}) reduces to
\be
( 1 - 4 \pi \chi_y - \tilde \kappa^2 ) 
\frac{\partial B_y}{\partial x}  = 
- \frac{ 4 \pi n_s e^2}{m_z c^2} A_z
\ee
Taking the $z$ derivative shows that
the penetration depth $\lambda_z$ is given by
\be
1 / \lambda_z^2 = \frac{ 4 \pi n_s e^2}{m_z c^2}  / 
( 1 - 4 \pi \chi_z - \tilde \kappa^2)
\label{lz}
\ee
Thus the $\kappa$ term only gives a correction to
the penetration depth proportional to $\tilde \kappa^2$.
Hence again there is no asymmetry
between the geometries where the samples occupy
$x > 0$ or $x < 0$.  Similar argument as in the last subsection
shows that there is a discontinuity in magnetic
field near the sample surface $\propto \tilde \kappa^2$.

\noindent{(C)} We finally consider a field along $\hat z$.
Without loss in generality, we take the sample to
occupy $y > 0$. Translational invariance
along $x$ and $z$ are respected and all quantities
depend only on $y$.  It can be shown easily
that the magnetic field obeys the usual screening
equations and thus 
$B_z(y) = B_z(0) e^{ -y / \lambda_x}$ where
$1 / \lambda_x^2 = \frac{ 4 \pi n_s e^2}{m_x c^2}/( 1 - 4 \pi \chi_z)$
is the same as that in section (A) 
[except that the small correction due to spin susceptibility is 
  here now $ (1 - 4 \pi \chi_z)$
instead of $ (1 - 4 \pi \chi_y )$ in eq (\ref{lx})].
The peculiar feature here, however, is that from eq (\ref{My})
that $M_y \ne 0$ since $q_x \ne 0$.
One easily finds 
$M_y(y) = \frac{\tilde \kappa}{4 \pi}  B_z(0) e^{ - y / \lambda_x}$.
Thus there is a magnetization towards (if $\kappa > 0$)
or away from (if $\kappa < 0$) the inside of the sample
if $B_{z {\rm ext}} > 0$.


\section{Field Distribution of a Single Vortex}\label{sec:vortex}

\noindent{(A)} Now we study the magnetic field profile around a vortex
for magnetic field in the basal plane, chosen to be again
along $\hat y$.
For simplicity, we shall ignore the small spin susceptibilities $\chi_y$ in
the present section.
The basic equations are again eq (\ref{mssz}) and (\ref{mssx}),
with $M_y$, $J_x$, $J_z$ given by eq (\ref{My}),(\ref{Jx}),(\ref{Jz}).
In constrast to section \ref{sec:meissner}A,
 the vector potential must
depend both on $x$ and $z$, and thus
\be
B_y = \frac{\partial A_x}{\partial z} - 
   \frac{\partial A_z}{\partial x}
\label{By}
\ee
Due to the presence of these two contributions to $B_y$, the
terms linear in $\kappa$ does {\it not}
drop out  in eq (\ref{mssz})
(or equivalently eq (\ref{pBpz}) below).
Therefore it is not a priori obvious that the vortex field distribution will
obey $z \to - z$ symmetry.  We shall however show below
that this symmetry is respected at least to order $\tilde \kappa^2$.

We begin by performing a rescaling of coordinates by the
penetration depths, thus we write
\ba
z &=& \lambda_x \tilde z \\
x &=& \lambda_z \tilde x
\ea
It is convenient also to rescale the components of ${\bf A}$:
\ba
A_x &=& \lambda_x \tilde A_x \\
A_z &=& \lambda_z \tilde A_z
\ea
so that
\be
B_y = \frac{\partial \tilde A_x}{\partial \tilde z} - 
   \frac{\partial \tilde A_z}{\partial \tilde x}
\label{Bt}
\ee
In these variables, eq (\ref{mssz}) and eq (\ref{mssx}) become
\ba
\frac{\partial B_y}{\partial \tilde z} 
   -  \left( \tilde A_x + \frac{\tilde \Phi_0}{2 \pi}
                   \frac{\partial \phi}{\partial \tilde x}  \right)
  &=&
  \tilde \kappa \frac{\partial}{\partial \tilde z}
              \left( \tilde A_x + \frac{\tilde \Phi_0}{2 \pi}
                   \frac{\partial \phi}{\partial \tilde x} \right)        
      - \tilde \kappa B_y
           \label{pBpz}     \\
\frac{\partial B_y}{\partial \tilde x} 
   +  \left( \tilde A_z + \frac{\tilde \Phi_0}{2 \pi}
                   \frac{\partial \phi}{\partial \tilde z}  \right)
  &=&
  \tilde \kappa \frac{\partial}{\partial \tilde x}
              \left( \tilde A_x + \frac{\tilde \Phi_0}{2 \pi}
                   \frac{\partial \phi}{\partial \tilde x} \right)        
      + \tilde \kappa^2
     \left( \tilde A_z + \frac{\tilde \Phi_0}{2 \pi}
                   \frac{\partial \phi}{\partial \tilde z}  \right)
\label{pBpx}
\ea    
where $\tilde \Phi_0 \equiv \frac{\pi \hbar c}{e} 
     \frac{1}{\lambda_x \lambda_z}$ is a scaled
flux quanta (magnetic field).  Note that as usual,
 ${\bf B}$ and ${\bf J}$ must
vanish at large distances and the total flux is therefore given
by the flux quanta $\frac{\pi \hbar c}{e}$.  In our
scaled variables this condition becomes
$\int B \ d \tilde x d \tilde z = \tilde \Phi_0$.

It seems difficult to solve eq (\ref{pBpz}) and (\ref{pBpx})
for general $\tilde \kappa$.
We shall thus make use of the smallness of $\tilde \kappa$ to
solve these equations order by order in this parameter.
(Strictly speaking the penetration depth $\lambda_z$ already
contains a $\tilde \kappa^2$ correction, but the present
rescaling using this corrected $\lambda_z$ 
simplifies the calculations below substantially).

In the lowest (zeroth) order, we can drop all terms on
the right hand sides of eq (\ref{pBpz}) and (\ref{pBpx}).
The resulting equations are the standard equations
for the vortex.  For a vortex with positive flux along $y$,
we need to choose $\phi = - \theta$ and the solutions
are
\be
B_y^{(0)}  = \frac{\tilde \Phi_0}{2 \pi} K_0 (r)
\label{b0}
\ee
\be
{\bf A}^{(0)} = A^{(0)}_{\theta} \hat \theta
\label{a0}
\ee
where
\be
A_{\theta}^{(0)} = A_{\theta}^{(0)} (r) 
  =  \frac{\tilde \Phi_0}{2 \pi} 
      \left(  \frac{1}{r} - K_1(r) \right)
\ee
and we have introduced the cylinderical coordinates for
the {\it scaled} variables:  $r \equiv ( \tilde z^2 + \tilde x^2)^{1/2}$,
${\rm cos} \theta = \tilde z / r$, ${\rm sin} \theta = \tilde x / r$,
$\hat \theta = -{\rm sin} \theta \hat z + {\rm cos} \theta \hat x$.
$K_0$, $K_1$ are the modified Bessel functions.

To first order, we have, from eq (\ref{pBpz}) and (\ref{pBpx}),
\ba
\frac{\partial B_y^{(1)}}{\partial \tilde z} 
   -  \tilde A_x^{(1)} 
  &=&
  \tilde \kappa \frac{\partial}{\partial \tilde z}
              \left( \tilde A_x^{(0)} + \frac{\tilde \Phi_0}{2 \pi}
                   \frac{\partial \phi}{\partial \tilde x} \right)              
      - \tilde \kappa B_y^{(0)}
           \label{pBpz1}     \\
\frac{\partial B_y^{(1)}}{\partial \tilde x} 
   +  \tilde A_z^{(1)}
  &=&
  \tilde \kappa \frac{\partial}{\partial \tilde x}
              \left( \tilde A_x^{(0)} + \frac{\tilde \Phi_0}{2 \pi}
                   \frac{\partial \phi}{\partial \tilde x} \right)        
\label{pBpx1}
\ea    

We claim that $B_y^{(1)} = 0$.  Assuming this, using
eq (\ref{b0}), (\ref{a0}) and therefore
$  \tilde A_x^{(0)} + \frac{\tilde \Phi_0}{2 \pi}
                   \frac{\partial \phi}{\partial \tilde x}
 = -    \frac{\tilde \Phi_0}{2 \pi} 
 K_1 (r) {\rm cos} \theta $,
eq (\ref{pBpz1}) and (\ref{pBpx1}) become
\ba
\tilde A_x^{(1)} &=&  
\tilde \kappa \frac{\tilde \Phi_0}{2 \pi} 
\left[ K_0 (r) +  \frac{\partial}{\partial \tilde z}
           \left(  K_1 (r) {\rm cos} \theta \right) \right]   
  \label{a1x} \\
\tilde A_z^{(1)} &=&  
- \tilde \kappa \frac{\tilde \Phi_0}{2 \pi} 
\left[ \frac{\partial}{\partial \tilde x}
           \left(  K_1 (r) {\rm cos} \theta \right) \right]
\label{a1z}
\ea
Indeed, substituting these equations into (\ref{By}),
we get
\be
B_y^{(1)} = \tilde \kappa \frac{\tilde \Phi_0}{2 \pi}
   \left[  \left( \frac{\partial}{\partial \tilde z} K_0 (r) \right)
   +  \left( \frac{\partial^2}{\partial \tilde z^2}  +
           \frac{\partial^2}{\partial \tilde x^2} \right)
        \left(  K_1 (r) {\rm cos} \theta \right) 
       \right]
\ee
However, by properties of the Bessel function,
 $\left( \frac{\partial^2}{\partial \tilde z^2}  +
           \frac{\partial^2}{\partial \tilde x^2} \right)
        \left(  K_1 (r) {\rm cos} \theta \right) 
       = K_1 (r) {\rm cos} \theta $.  Further
using $K'_0 = - K_1$ shows that indeed $B_y^{(1)} = 0$.

Now we proceed to the second order.  Eq (\ref{pBpz}) and 
(\ref{pBpx}) read
\ba
\frac{\partial B_y^{(2)}}{\partial \tilde z} 
   -  \tilde A_x^{(2)} 
  &=&
  \tilde \kappa \frac{\partial}{\partial \tilde z}
               \tilde A_x^{(1)}           \label{pBpz2}     \\
\frac{\partial B_y^{(2)}}{\partial \tilde x} 
   +  \tilde A_z^{(2)}
  &=&
  \tilde \kappa \frac{\partial}{\partial \tilde x}
              \tilde A_x^{(1)} 
            + \kappa^2  
       \left(     \tilde A_z^{(0)} + \frac{\tilde \Phi_0}{2 \pi}
                   \frac{\partial \phi}{\partial \tilde z}  \right)        
\label{pBpx2}
\ea  
We again claim that $B_y^{(2)} = 0$.  If so,
we get, using
$\tilde A_z^{(0)} + \frac{\tilde \Phi_0}{2 \pi}
                   \frac{\partial \phi}{\partial \tilde z} $
 $ = $  
$\frac{\tilde \Phi_0}{2 \pi}  K_1 (r) {\rm sin } \theta $, 
and eq (\ref{a1x}) and (\ref{a1z}),
\ba
\tilde A_x^{(2)} &=&  
- \tilde \kappa^2 \frac{\tilde \Phi_0}{2 \pi} 
\left[  \frac{\partial K_0 (r) }{\partial \tilde z} +  
   \frac{\partial^2}{\partial \tilde z^2}
           \left(  K_1 (r) {\rm cos} \theta \right) \right]   
  \label{a2x} \\
\tilde A_z^{(2)} &=&  
 \tilde \kappa^2 \frac{\tilde \Phi_0}{2 \pi} 
\left[ \frac{\partial K_0 (r) }{\partial \tilde x} +
          \frac{\partial^2}{\partial \tilde x \partial \tilde z}
     \left(  K_1 (r) {\rm cos} \theta \right) 
            + K_1 (r) {\rm sin} \theta  \right]
\label{a2z}
\ea
From eq (\ref{By}) for $B_y^{(2)}$,
we get
\be
B_y^{(2)} = - \tilde \kappa^2 \frac{\tilde \Phi_0}{2 \pi}
   \left[  \tilde \nabla^2 K_0 (r) 
   +  \frac{\partial}{\partial \tilde z} \tilde \nabla^2
         \left( K_1 (r) {\rm cos} \theta \right)  +
           \frac{\partial}{\partial \tilde x} 
        \left(  K_1 (r) {\rm sin} \theta \right) 
       \right]
\ee
Using again the properties of the modified Bessel functions,
this reduces
to
\be
B_y^{(2)} = - \tilde \kappa^2 \frac{\tilde \Phi_0}{2 \pi}
   \left[   K_0 (r) 
   +  \frac{\partial}{\partial \tilde z} 
         \left( K_1 (r) {\rm cos} \theta \right)  +
           \frac{\partial}{\partial \tilde x} 
        \left(  K_1 (r) {\rm sin} \theta \right) 
       \right]
\ee
The last two term combines to 
$ \frac{1}{r} \frac{d}{dr} \left( r K_1 (r) \right) = - K_0$
by recursion relation of modified Bessel function.  Hence
$B_y^{(2)} = 0$ as claimed.

Hence to second order in $\tilde \kappa$, 
$B_y$ has the same form as an ordinary anisotropic
material:
$B_y = \frac{\Phi_0}{ 2 \pi \lambda_x \lambda_z} K_0 
\left(  \left[ (z/\lambda_x)^2 + (x/\lambda_z)^2 \right]^{1/2} \right)$.
The constant magnetic field contours are ellipses
with center at the point where the order parameter has
a singularity.

We would however like to add two cautionary remarks.
Firstly,  it is
{\it not} true that other physical quantities 
such as ${\bf J}$ or ${\bf M}$ also
have elliptic distributions around the vortex.
For example, using eq (\ref{My}) and the solution
to ${\bf A}^{(0)}$, we find
$M_y^{(1)} = -\tilde \kappa  \frac{\tilde \Phi_0}{ 8 \pi^2}
  K_1(r) {\rm cos} \theta $.
Therefore $M_y(x,z)$ is odd under $z \to -z$.
Secondly, it is not true that
the corrections to magnetic field vanish
to higher orders.  Proceeding to the third order,
 one can verify that $B_y^{(3)} = 0$ is
{\it inconsistent} with ${\bf A}^{(2)}$ of eq (\ref{a2z})
and (\ref{a2x}).  We, however would not proceed to
calculate these very small corrections.

\noindent{(B)} We now consider a vortex with field
along $\hat z$.  Using translational invariance along $z$,
one finds that there is no magnetic field induced
along the $x-y$ plane, and $B_z$ and ${\bf A}$ are
the same as 
those of an ordinary superconductor with
in-plane penetration depth $\lambda_x$ given 
in Sec \ref{sec:meissner} before.
However, due to the presence of ${\bf q}$,
there is an in-plane magnetization 
induced by the presence of the $\kappa$ term  
(see eq (\ref{Mx}) and (\ref{My})).
We find that this magnetization is along the radial
direction:  ${\bf M} = M_r \hat r$ 
(thus its curl vanishes) where
$M_r =  \tilde \kappa  \frac{\Phi_0}{ 8 \pi^2 \lambda_x^2} 
  K_1 \left( \frac{(x^2+y^2)^{1/2}} {\lambda_x} \right)$.
The magnitude of this magnetization is therefore
$\tilde \kappa / 4 \pi$ times the local magetic
field $B_z$ along the $z$ direction.  It points radially
outwards if $\kappa > 0$.  For field along $-\hat z$,
this magnetization changes sign and points radially
{\it inwards} if $\kappa > 0$.


\section{Order of Magnitude}\label{sec:es}

Finally we estimate $\tilde \kappa$, assuming the clean limit.
The calculations in Ref \cite{Yip02} can be easily
generalized to the present 3D case once the Fermi surface
and the spin-orbit splittings are given.  We however
would not do this calculation here but just satisfy ourselves
with some estimates.  It should be noted that,
for the crystal symmetry $P4mm$ in question, the allowed
spin-orbit interaction, besides one in the Rashba form
$ - \alpha \hat z \cdot (\vec p \times \sigma)$ $=$
$ - \alpha (p_x \sigma_y - p_y \sigma_x)$,
(here ${\bf p}$ is the momentum, ${\bf \sigma}$ the Pauli matrix,
and $\alpha$ a coefficient)
can also have terms of the form
$ - \beta  p_x p_y ( p_x \sigma_x - p_y \sigma_y)$
and $ - \gamma p_x p_y p_z ( p_x^2 - p_y^2 ) \sigma_z$.
Here $\alpha$, $\beta$, $\gamma$ can be functions of ${\bf p}$
but must be invariant under the symmetries of the crystal.
(c.f., \cite{Yip93}).
Both the first two terms can generate 
the terms proportional to $\kappa$ in eq (\ref{Jx})-(\ref{My}).
However, we expect that (c.f., eq (12) of ref \cite{Yip02})
that $\kappa$ (at $T \to 0$) has an order of magnitude
given by
$\sim \mu_{B} ( p_{F+}^2 - p_{F-}^2 ) / \hbar^3 $  where
$\mu_B$ is the Bohr magneton, $p_{F \pm}$ are the 
typical Fermi momenta for the spin-orbit splitted Fermi surfaces.
Using the expression for $\lambda_x$ and $n_s \sim p_F^3$,
we find that $\tilde \kappa$ is of order
\be
\tilde \kappa \sim 
 \left( \frac{e^2}{\hbar c} \right)
 \left( p_F r_B  \right)^{1/2}
 \left( \frac{\delta}{\mu} \right)
\ee
where $\delta$ is the typical splitting in energy by the spin-orbit
interaction,
$\mu$ the chemical potential, and $r_B$ the Bohr radius.
For $\delta/\mu \sim 0.1$, this ratio is then of order
$\sim 10^{-3}$ assuming typical electron densities.


\section{Conclusion}

In conclusion, we have studied some magnetic properties
for a superconductor with no inversion symmetry in its
normal state.  In particular we investigated how the 
broken symmetry and magneto-electric effects
discussed in ref \cite{Yip02,Edelstein95} exhibit
themselves in Meissner screening and vortices.
An unusual magnetization spatial pattern
are found in some geometries.  This magnetization
can in principle be detected by Knight shift
measurements.

\section{Acknowledgement}

 This research was supported by National Science Council of Taiwan 
under grant numbers NSC
92-2112-M-001-041 and 93-2112-M-001-016.


\end{document}


\title{Errata
} \subtitle{ Magnetic Properties of a Superconductor with no
Inversion Symmetry, by S. K. Yip, J. Low Temp. Phys. 140, 67 (2005)}


\author{S. K. Yip
}


\institute{S. K. Yip \at
              Institute of Physics \\
              Academia Sinica\\
              Taiwan\\
              Tel.: +886-2-27896785\\
              Fax: +886-2-27834187\\
              \email{yip@phys.sinica.edu.tw}           
}

\date{Received: date / Accepted: date}

\maketitle



The claim that $B_y^{(1)} = 0$ in Sec 4A of the above paper
(hereafter as I) is incorrect.  The calculation over there only
establishes a particular solution Eqs. I(33), I(34) for $\tilde
A_{x,z}^{(1)}$ to Eqs. I(31), I(32) with $B_y^{(1)} = 0$.  It is
necessary to add a solution to the homogeneous equations
corresponding to Eqs. I(31) and I(32), {\it i.e.}, \ba
\frac{\partial B_y^{(1),h}}{\partial \tilde z}
   -  \tilde A_x^{(1),h}
  &=&
  0
           \label{h1}     \\
\frac{\partial B_y^{(1),h}}{\partial \tilde x}
   +  \tilde A_z^{(1),h}
  &=&
  0
\label{h2}
 \ea
 correspondingly
 \be
\frac{\partial^2 B_y^{(1),h}}{\partial \tilde z^2} +
\frac{\partial^2 B_y^{(1),h}}{\partial \tilde x^2}  - B_y^{(1),h} =
0  \ . \label{h3} \ee This homogeneous solution must be included so
that $B_y^{(1)} - 4 \pi M_y^{(1)} $ contains no discontinuity at the
vortex. Using $M_y^{(1)} = \frac{\kappa}{2} q_x^{(0)}$ hence \be
M_y^{(1)} = -\tilde \kappa \frac{\tilde \Phi_0}{ 8 \pi^2}
  K_1(r) {\rm cos} \theta \ ,
  \label{My}
  \ee
  the homogeneous solution required is
  \be
  B_y^{(1),h} =
-\tilde \kappa  \frac{\tilde \Phi_0}{ 2 \pi}
  K_1(r) {\rm cos} \theta
  \label{byh}
  \ee
  correspondingly $B_y^{(1)} = B_y^{(1),h}$.  We see now that
  $B_y^{(1)} - 4 \pi M_y^{(1)}$ becomes identically zero.  To check
  the consistency with the Maxwell equation I(8), we note that
  $ \tilde A_{x,z}^{(1),h} $ can be obtained from Eqs (\ref{h1}) and
  (\ref{h2}).  Together with Eqs. I(33) and I(34), we obtained the
  correct first order vector potentials as
  \be
  \tilde A_x^{(1)} = \tilde \kappa \frac{\tilde \Phi_0}{ 2 \pi}
  K_0(r)
  \label{Ax}
  \ee
  and
  \be
 \tilde A_z^{(1)} = 0
 \label{Az}
 \ee
 Eq (\ref{Az}) gives $J_z^{(1)} = 0$.  Eq (\ref{Ax}), together with
 I(1) and $B_y^{(0)} = \frac{\tilde \Phi_0}{ 2 \pi}
  K_0(r)$ gives $J_x^{(1)} = 0$.

  The total field up to first order is therefore
  \ba
  B_y^{(0)} + B_y^{(1)} &=& \frac{\tilde \Phi_0}{ 2 \pi}
  \left[ K_0(r)  - {\tilde \kappa} K_1(r) {\rm cos} \theta \right]
  \\
 &\approx&
\frac{\tilde \Phi_0}{ 2 \pi}
  \left[ K_0(|\vec r + {\tilde \kappa} \hat z |) \right]
  \label{btot}
  \ea

  Therefore, to first order in $\tilde \kappa$, the field pattern
  is the same as a centrosymmetric superconductor except that it
  is shifted from the vortex center towards $-z$ by $\tilde \kappa$
  in dimensionless units (${\tilde \kappa} \lambda_x$ in ordinary
  units).  This shift has been found numerically by Oka et al
  \cite{Oka}
  in their numerical solution of Ginzburg-Landau
  equations, instead of the London equations here.

  In contrast, the total gauge invariant superfluid velocities along $x$ and $z$ are, up to first order in
  $\tilde \kappa$, proportional to
  $-K_1(r) {\rm cos} \theta + {\tilde \kappa} K_0(r)$ and
  $K_1(r) {\rm sin} \theta$, respectively.  The singularity
  is dominated by $K_1(r)$ terms and is still at the vortex center.

  The correct $\tilde A_{x,z}^{(1)}$ in Eq (\ref{Ax}) and (\ref{Az})
  has to be used to obtain the next higher correction $B_y^{(2)}$,
  which can be shown to be finite instead of zero as claimed in I.
   The expression is not particularly informative and we shall not
   obtain it here.

  The results of the other sections in I are unaffected.


